\documentclass{appolb}
\usepackage{epsfig}
\begin{document}
\title{$\pi\omega\rho$ vertex in nuclear matter
%
%\thanks{Presented at Meson 2002, Krak\'ow}%
}
\author{Agnieszka Bieniek
\address{The H. Niewodnicza{\'n}ski Institute of Nuclear Physics, 
ul. Radzikowskiego 152 PL-31342 Cracow, Poland}}
%\and
%Anna Baran, Wojciech Broniowski
%\address{The H. Niewodnicza{\'n}ski Institute of Nuclear Physics, 
%ul. Radzikowskiego 152 PL-31342 Cracow, Poland}}
\maketitle

\begin{abstract}
{
Medium modifications of
the $\pi \omega \rho$ vertex are analyzed in context of the
$\omega \to \pi^0  \gamma^\ast$ and $\rho \to \pi \gamma^\ast$
decays in nuclear matter. A relativistic hadronic model 
with mesons, nucleons, and $\Delta(1232)$ isobars is applied. 
A substantial increase of the widths for the decays
$\omega \to \pi^0 \gamma^\ast$ and $\rho \to \pi \gamma^\ast$ is found 
for photon virtualities in the range $0.3-0.6$GeV. This enhancement has
a direct importance for the description of dilepton yields from dalitz decays
in relativistic heavy-ion collisions.}
\end{abstract}
\PACS{25.75.Dw;21.65.+f;14.40.-n}
%\bigskip
\vspace{0.1cm}
The experience gathered over the past years led to the conviction that
the properties of hadrons in medium are substantially
modified.
For instance, it has been found that the value of the 
$\rho \pi \pi $ coupling is considerably enhanced.
In Ref. \cite{aa}, which is the basis for 
this talk, 
we have examined the {\em %
in-medium} $\pi \omega \rho $ coupling. This vertex enters the 
decays $\omega \rightarrow \pi ^{0}\gamma ^{\ast }$ and 
$\rho \rightarrow \pi \gamma^{\ast }$, which 
are important for description of the dilepton production 
in relativistic heavy-ion collisions \cite{cereshelios}. 
In our study we use a typical hadronic model with meson, nucleon
and $\Delta$ degrees of freedom, and for simplicity 
work at leading baryon density and at zero
temperature.

In the vacuum the $\pi \omega \rho$ coupling is 
\begin{equation}
-iV_{\pi \omega^\mu \rho^\nu }=i\frac{g_{\pi \omega \rho }}{F_{\pi }}%
\epsilon ^{\mu \nu p Q},  \label{vac}
\end{equation}
where we use the notation $\epsilon ^{\mu \nu
pQ}=\epsilon ^{\mu \nu \alpha \beta }p_{\alpha }Q_{\beta }$, and
$F_\pi=93$MeV denotes the pion decay constant. In our convention 
$Q$ is the incoming momentum of the pion, $p$ is the outgoing momentum of
the $\rho$, and $q = Q-p$ is the outgoing momentum of the $\omega $. The vector
mesons are, in general, virtual.
The nucleon propagator in nuclear matter is decomposed into the {\em %
free}, $S_{F}$, and {\em density}, $S_{D}$, parts in the usual way. 
The Rarita-Schwinger propagator is used for the spin 3/2  $\Delta(1232)$ particle. 
The effects of the non-zero width of the $%
\Delta $ are incorporated by replacing 
$M_{\Delta }$ by the complex mass $M_{\Delta }-i\Gamma /2$.

\vspace{-1.2cm}
\begin{figure}[htb]
\centerline{\psfig
{figure=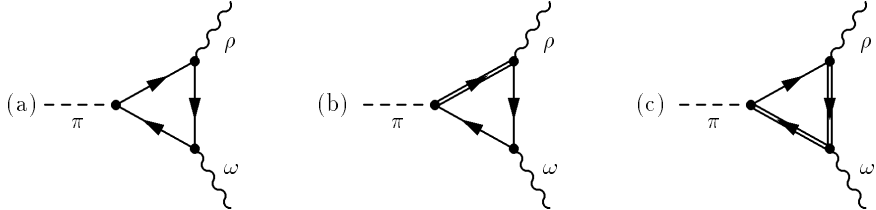,height=4.4cm,bbllx=75bp,bblly=377bp,bburx=541bp,bbury=537bp,clip=}}

%\vspace{-.5cm}

\label{diag}
\caption{Diagrams included in our calculation (crossed not shown).
Solid lines denote
the in-medium nucleon, and double lines the $\Delta$ resonance.}
\end{figure}

The diagrams used in our study are shown in Fig. 1.
Density effects are produced when exactly one of the nucleon lines 
in each of the our diagrams of Fig. 1 involves the nucleon 
{\em density} propagator, $S_{D}$. Details can be found in Ref.\cite{aa}.
We analyze the $\omega \rightarrow \pi ^{0}\gamma ^{\ast }$ and 
$\rho \rightarrow \pi \gamma^{\ast }$ decays where the $\omega$ 
and $\rho$ meson move with a non-zero momentum $\bf q$ with respect to the
medium. The decaying particle can have transverse or longitudinal 
polarization, defined by quantizing the spin along the direction of $\bf q$.
The properties of the $\omega$ and $\rho$ mesons are different 
for these two polarizations, in particular their widths $\Gamma^L$ and $\Gamma^T$
are different. We observe
that the medium effect weaken with increasing $|\bf q|$. However, they remain large
for $|\bf q|$ lower than 200~MeV, the relevant values for heavy-ion collisions.

The theoretical output
from the Dalitz decays of vector mesons becomes enhanced in the region
of $0.3-0.6$GeV, which is exactly where the existing calculations have
serious problems in supplying enough strength to explain the CERES and HELIOS
data \cite{cereshelios}. The dilepton production rate from the Dalitz
decay of a vector meson is given by the formula \cite{PKoch} 
\begin{equation}
\frac{dN_{l^{+}l^{-}}}{d^{4}x\,dM^{2}}=\int \frac{d^{3}p_{v}}{(2\pi
)^{3}E_{v}}\,f(p_{v })\frac{m_{v }}{\pi M^{3}}(2\Gamma
_{v \rightarrow \pi ^{0}\gamma ^{\ast }}^T+\Gamma
_{v \rightarrow \pi ^{0}\gamma ^{\ast }}^L)\,\Gamma _{\gamma ^{\ast
}\rightarrow l^{+}l^{-}}  \label{PK}
\end{equation}
where $v$ labels the vector meson, and $f(p_v)=exp\left(\frac{-E_v}{T}\right)$ is the  
distribution function.
The width $\Gamma _{v \rightarrow \pi ^{0}\gamma ^{\ast }}$ depends on
the baryon density, $\rho _{B}$, which in turn depends on the space-time
point $x$. Clearly,  an increased value of
$\Gamma _{v \rightarrow \pi ^{0}\gamma ^{\ast }}$
results in an increased dilepton yield.
A more exact estimate requires a detailed model of the evolution
of the fireball. In order to calculate the dilepton spectrum we adopt the model
of the fire cylinder from Ref. \cite{RappWamb}.
%\emph {R. Rapp, J. Wambach, Eur.Phys.J.A 6 (1999)415.}
The evolution model includes longitudinal and transverse expansion, and cooling. 

\begin{figure}[htb]
\centerline{
%\vspace{0mm} ~\hspace{-2.5cm} 
%\epsfxsize = 12cm \centerline{\epsfbox{omrho11.eps}} \vspace{0mm}
\epsfysize = 6cm \centerline{\epsfbox{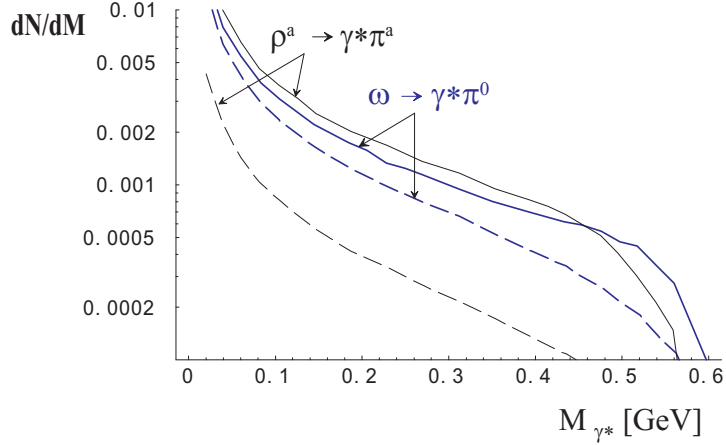}} \vspace{0mm}} 
%\includegraphicx[width = 0.6\linewidth]{omrho11.eps}
\label{fig:2}
\caption{The dilepton yield  for the decay $%
\omega \to \gamma^\ast \pi^0$ (thick lines) and $\rho^a \to \gamma^\ast \pi^a$ (thin lines), 
plotted as a function of the virtual mass of $\gamma^\ast$, $M$. Dashed and solid lines 
correspond to the vacuum and in-medium cases, respectively.}
\end{figure}

%\eject
\noindent The dilepton production rate is given by
\bigskip
\begin{eqnarray}
\frac{dN}{dM} = \int\limits_0^{t_{max}} dt \int\limits_0^{r_{max} (t)} 2\pi r dr \int\limits_{-z_{max}
(t_{max})}^{z_{max}(t_{max})} dz \left( \frac {dN}{d^4 x dM}\right)  
\end{eqnarray}

Our numerical results of the Dalitz decays ($\omega \to \gamma^\ast \pi^0$ and $\rho^a \to \gamma^\ast \pi^a$)
are shown in Fig. 2. 
In both cases the medium effects
are very important: they bring an enhancement by a factor of $\sim 3$ for the $\omega$ and by a factor of
$\sim 5$ for the $\rho$. We wish to point out that the Dalitz decays of the $\rho$ meson should not 
be neglected, as is frequently done.

This research has been done together with W. Broniowski and 
partly with A. Baran.

\end{document}